\documentclass{article}
\usepackage{arxiv}
\usepackage[T1]{fontenc}    
\usepackage{hyperref}       
\usepackage{url}            
\usepackage{booktabs}       
\usepackage{amsfonts}       
\usepackage{nicefrac}       
\usepackage{microtype}      
\usepackage{lipsum}
\usepackage{algpseudocode}
\usepackage{amsmath}
\usepackage{float}
\usepackage{amsmath,amssymb,amsfonts}
\usepackage{multirow}
\usepackage{algorithm}
\usepackage{graphicx}
\usepackage{xcolor}
\usepackage{setspace}
\usepackage[style=ieee,backend=biber]{biblatex}
\graphicspath{ {./images/} }
\title{Predict future sale}
\title{FastGEMF: Scalable High-Speed Simulation of Stochastic Spreading Processes over Complex Multilayer Networks}

\author{
  Mohammad Hossein Samaei \\
  Kansas State University\\
  Manhattan, KS, USA \\
  \texttt{msamaei@ksu.edu} \\
   \And
 Faryad Darabi Sahneh \\
  University of Arizona\\
  Tucson, AZ, USA\\\
  \And
 Caterina Scoglio \\
  Kansas State University\\
   Manhattan, KS, USA \\
}
\begin{document}
\maketitle
\begin{abstract}
Predicting the spread of processes across complex multi-layered networks has long challenged researchers due to the intricate interplay between network structure and propagation dynamics. Each layer of these networks possesses unique characteristics, further complicating analysis.To authors' knowledge, a comprehensive framework capable of simulating various spreading processes across different layers, particularly in networks with millions of nodes and connections, has been notably absent. This study introduces a novel framework that efficiently predicts Markov Chain processes over large-scale networks,  while significantly reducing time and space complexity. This approach enables exact simulation of spreading processes across extensive real-world multi-layer networks, accounting for diverse influencers on each layer. FastGEMF provides a baseline framework for exact simulating stochastic spread processes, facilitating comparative analysis of models across diverse domains, from epidemiology to social media dynamics.
\end{abstract}
\label{sec:introduction}
Spreading  phenomena, observed across diverse networks, involve processes like the diffusion of digital viruses, memes, and infectious diseases such as COVID-19. Nodes in these networks represent various entities—from humans to devices and are linked through edges, facilitating dynamic interaction models. Pioneering work in epidemic modeling introduced models like SIS and SIR\cite{kermack1927contribution}, which have been fundamental in understanding these processes and their impact on human societies \cite{Daley_Gani_1999} With the rise of the digital era, similar analytical approaches have been applied to information technology, exploring the spread of rumors \cite{10.1145/2184319.2184338}, misinformation \cite{10.1145/2503792.2503797}, and societal opinions \cite{watts2007influentials}. Additionally, financial networks and cybersecurity are increasingly studied for their propagation dynamics, utilizing epidemiological methods for broader applicability \cite{gai2010contagion,kanno2015assessing, barro2010credit, meng2018securing, cebe2018network}.

Epidemic spreading dynamics on networks evolve from simple homogeneous networks, characterized by classical models such as complete graphs, to more realistic scale-free and random networks\cite{Barrat_Barthélemy_Vespignani_2008} represented  by the Barabási-Albert \cite{barabasi1999emergence}, Erdős-Rényi \cite{erdds1959random}, and Watts-Stogatz \cite{watts1998collective} models. Initial studies focused on single-agent spread over single-layer networks, highlighting the impact of human behaviors and information dissemination \cite{funk2010modelling, funk2015nine, kwon2013prominent}. Simulators such as Epidmeic on Networks\footnote{EoN: https://bit.ly/EoNGit} (EoN) \cite{kiss2017mathematics}, Nepidemix\footnote{Nepidemix: https://goo.gl/M8rEGM}, and Epydemic\footnote{Epydemic: https://goo.gl/PrPHh4} are among those developed for single agent models, e.g. SIR and SIS models over one layer networks.For instance, Kiss et al. \cite{kiss2017mathematics}, presented two specialized  modules as \textit{$Fast\_SIR$} and \textit{$Fast\_SIS$} for scalable simulation over large single-layer networks.  Furthermore, the complexity increased as more agents were employed in mechanistic models, known as multi-agent mechanistic models, in which different influencers compete with each other\cite{karrer2011competing,funk2009spread, newman2005threshold,ruan2012epidemic,zhan2018coupling}. Two famous modules for multi-agent spread over network are EpiModel\footnote{EpiModel: https://goo.gl/PrPHh4} in R language \cite{jenness2018epimodel} and Network Diffusion library \footnote{NDlib: https://bit.ly/NDlibGit} (NDlib) in Python language \cite{rossetti2018ndlib}, while both are able to run multi-agent models over both temporal and static networks. Specifically, EpiModel was designed for modeling temporal networks, usually in sizes of less than tens of thousands, to accurately simulate the structure of networks for small communities, with focus on sexually transmitted diseases (STDs). Subsequently, the focus shifted to single spread across multi-layer networks, accommodating different relational types like physical contacts and social media, with the added intricacies of weighted and directed edges \cite{cozzo2013contact, sun2014epidemic}. This paved the way for more complex models in multi-layer networks (known as networks with the same nodes on different layers and different links, sometimes also referred to as multiplex networks), exploring the simultaneous impacts of various agents, from epidemiology to financial markets. \cite{cozzo2013contact, sun2014epidemic, de2016physics, d2014modeling}.In \cite{sahneh2017gemfsim}, Sahneh et al.  introduced a novel general framework for prediction of spread process over multi-layer networks and completely investigate the complexity and feasibility of finding analytical solution. To tackle the state explosion phenomena in Markov Chain process, they introduced a Generalized Mean Field Model, which tackle the infeasibility of analytical solution using mean-field approximation. To investigate the accuracy of this approximation, they presented a  module for exact simulation of spread over stochastic processes over multi-layer networks  as Generalized Epidemic Modeling Framework simulator (GEMFsim) in 4 different languages, MATLAB, R, C, and Python. The framework's generality and exactness, while advantageous, incurred high computational costs, limiting its scalability to large networks up tens of thousands of nodes.  For further research on multi-agent spread over multilayer networks, refer to  \cite{ poledna2015multi, jankowski2016picture,lin2023epidemic, erlandsson2018seed,peng2021multilayer, salehi2015spreading, darabi2014competitive,sood2023spreading}, and For a comprehensive systematic review of the topic, see  \cite{brodka2020interacting}.

Surprisingly, despite significant advancements in this field, there remains a notable gap in frameworks capable of scaling to networks comprising millions of nodes. Most studies have concentrated on networks ranging from 100 to a maximum of 10,000 nodes—scales that fall short of addressing many real-world scenarios \cite{kuryliak2024efficient, jenness2018epimodel}. A few researches have addressed scalability to large scale networks, employing aggregation techniques, specialization methods, or massive parallel computing \cite{gomez2023new,Kiss2017,bianconi2018multilayer, doi:10.1177/0037549711413001}. These approaches enable handling larger networks but often involve simplifying network structure, limiting to specific type of spread, or requiring significant computational resources.\\

A few researchers have developed agent-based model frameworks like EpiHiper \cite{bhattacharya2023data}, Loimos \cite{kitson2024large}, and EpiSimdemics \cite{bhatele2017massively}, tested on supercomputers (Rivanna, Blacklight, Bridges2, Blue Waters), which require substantial computing resources that are not readily accessible to most researchers and practitioners in the field.\\

In this work, we have developed a novel algorithm that is both scalable and capable of simulating any  multi-compartment stochastic spreading processes across multiple layers of complex networks. This work builds on the foundation of the GEMF  and inspired by exact algorithm proposed by Gillespie\cite{gillespie1976general, gillespie1977exact}  and later optimized by Gibson et al. \cite{gibson2000efficient} for chemical reactions, to include any competitive-like spreading process that exhibits Markov properties (refer to \cite{karlin1981second} for details). The proposed framework  does not  resort to any approximations allowing us to maintain accuracy and generality while efficiently handling the stochastic nature of the spread.
Our algorithm has been rigorously tested on both dense and sparse networks, and across multilayer networks with different topologies, including real-world networks, scale-free networks, and others. It can handle directed or undirected, as well as weighted or unweighted scenarios. Furthermore, the software developed from this study is provided open source for all users. The main contributions of this paper are summarized as follows:\\
\begin{itemize}
\item Introduction of a novel, optimized algorithm that leverages the Gillespie algorithm within the Generalized  Epidemic Modelling Framework  (GEMF), enhancing the simulation of epidemic processes across complex network structures.
\item Optimization of the package to achieve low time-complexity as \( \textbf{O}(D_{max}\log(N))\), while $N$ and $D_{max}$ are the number of nodes and maximum node degree, respectively. Ensuring efficient processing and scalability  as network size increases.
\item Provision of the software as open-source module in Python, facilitating accessibility and contribution for both the general public and researchers. This allows for widespread use and further development by the academic community.
\end{itemize}

\section{Methodology}

In this section, we first explore the fundamentals of  GEMF, discussing its basic mathematical foundation. We then examine the inefficiencies of traditional Gillespie-based algorithms and GEMFsim, and present our approach for enhancing their efficiency. Finally, we provide a detailed description of how the proposed algorithm is implemented, ensuring a comprehensive understanding of its operational mechanics.
\subsection{The Framework for Stochastic Multispreading Over Multilayer Networks: GEMF}
\begin{figure}[H]
	\centering
	\includegraphics[width=1\linewidth]{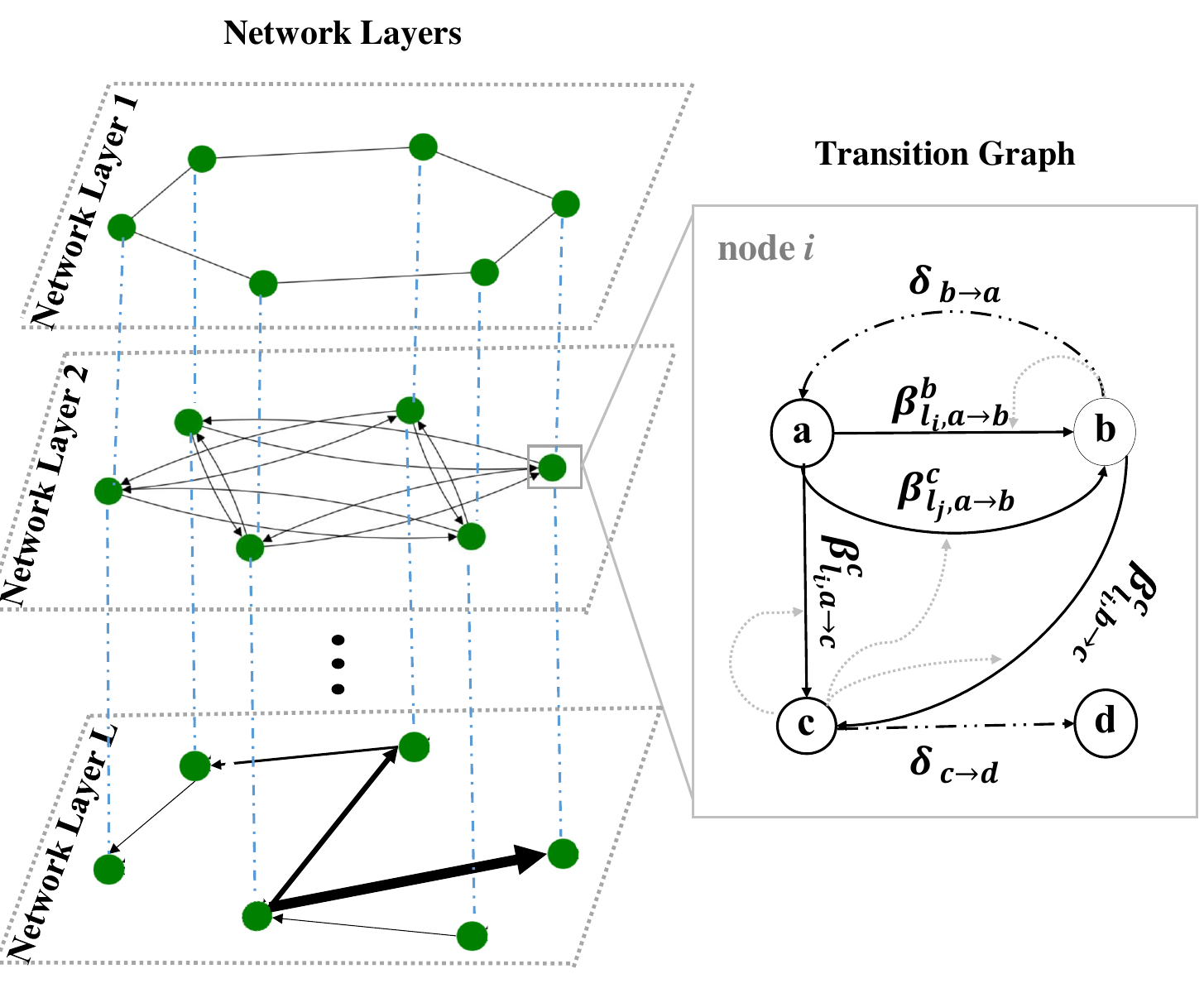}
	\caption{Overall Structure  complex multiplayer networks along with multi-agent transition graph that each node undergoes to. In the contact graphs, the solid line, arrows and width of lines denotes undirected, directed and weighted properties of the graphs, respectively. For the transition graphs solid and dashed arrows show edge-based and node-based transitions, respectively. Dotted grey arrows specify the state which induces the transition. \(\beta^c_{l_i,a\rightarrow b}\) denotes the rate for transition from state $a$ to $b$ under 
  interaction with neighbors which are in state $c$ at layer $l_i$. $\delta_{b\rightarrow a}$ is for the rate of transtition from state $b$ to $a$, which is independent of its interaction with other individuals  } 
\end{figure}

In \cite{sahneh2013generalized}, Sahneh et al. proposed a structure to facilitate the study of a wide range of spreading processes over complex contact networks, featuring multiple layers and a flexible structure. Figure 1 illustrates the overall idea behind representing multilayer\footnote{ These layers are also referred to as multiplex networks in the literature, where each layer represents a different type of connection or interaction between nodes} graph topology as \(\mathcal{G}(\mathcal{N},E_1,E_2,\ldots,E_L)\), where \(\mathcal{N}\) represents the nodes and \(E\) represents the different layers. Layers can be representative of different types of interaction; for example, one layer can represent a physical contact network, a second layer an information network, and a third layer a sexual network. Different structures of networks are shown in this diagram to emphasize that in the proposed framework, no assumptions are made about the structure of complex networks across different layers.The transition graph in Figure 1 illustrates multi-compartment mechanistic models that show how the state of each node (e.g., node $i$) is governed by dynamical interactions between individuals in the network. For example, from the transition model in Figure 1 we can see following interactions:
\begin{enumerate}
    \item The transition of node $i$ from state $b$ to state $c$ results from interaction with its neighbors at layer $l_i$ being in state $c$. We assume that the probability of node $i$ remaining in state $b$ decays exponentially with rate $\beta^c_{l_i, b\rightarrow c}$.
    \item When node $i$ is in state $a$, it can change to state $b$ through interaction with  two different influencers at two distinct layers:
    \begin{itemize}
        \item Individual neighbors at layer $l_i$ in state $c$, with rate $\beta^c_{l_j, a\rightarrow b}$
        \item Individual neighbors at layer $l_j$ in state $b$, with rate $\beta^b_{l_i, a\rightarrow b}$
    \end{itemize}
    \item Individual neighbors in states $c$ and $b$ can compete to infect node $i$ when it's in state $a$, with rates $\beta^c_{l_i, a\rightarrow c}$ and $\beta^b_{l_i, a\rightarrow b}$ respectively.
    \item Some transitions occur without interaction with other individuals, such as the transition from state $b$ to state $a$ with rate $\delta_{b \rightarrow a}$, similar to the curing process in the SIS model.
\end{enumerate}
The only assumption in these mechanistic models is the independence of competing processes \cite{sahneh2013generalized}.\\
\textbf{Assumption 1}: A transition from state \( a \) to state \( b \) for agent \( i \) results from multiple stochastically independent competing processes. The transition \( a \to b \) for agent \( i \) occurs either independently of the states of other agents or due to interactions between agent \( i \) and agent \( j \), where \( j \in \{1, \dots, N\} \setminus \{i\} \).\\
Therefore, each layer can have different or identical inducers, and each inducer can cause transitions to different states at different rates, and according to Assumption 1, the interaction of node $i$ with node  $j \ne i$ is stochastically independent of its interaction with node $k \notin \{i, j\}$.  This presentation emphasizes on two types of agent-based transitions that are defined in these model categories:

\begin{itemize}
    \item \textbf{Node-based transition} - describes a type of transition independent of the contact network (e.g., the curing process in the SIS model),
    \item \textbf{Edge-based transition} - results from the interactions between agents within a network (e.g., the infection process in the SIS model).
\end{itemize}

Considering \(x_i(t) \in \{e_1, \dots, e_M\}\) as the state of the \(i\)th agent at a given time \(t\), with \(e_m\) being the \(m\)th standard unit vector in \(\mathbb{R}^{M}\) (e.g., in the SIS model, \(e_1 = \begin{bmatrix} 1 \\ 0 \end{bmatrix}\) and \(e_2 = \begin{bmatrix} 0 \\ 1 \end{bmatrix}\) represent the agent in the "susceptible" and "infected" states, respectively). Therefore, the network state as the joint state of all agents can be defined as:
\begin{equation}
    X(t)=[x_1(t),\dots, x_n(t)] 
\end{equation}
Where \( X \in \mathbb{R}^{M \times N} \) represents the continuous time Markov chain (CTMC) with \( M^N \) different possible outcomes. To find the dynamics of the network state as it evolves during the time interval \( \Delta t \) for a possible transition \( m \rightarrow n \), conditioned on \( X(t) \) with \( m \) being the current state of the \( i \)th node, we have:

\begin{equation}
\begin{split}
\Pr[x_i(t + \Delta t) &= e_n \mid x_i(t) = e_m, X(t)] \\
&= \sum_{j=1}^{N} I_{(i,j)}(t) \cdot T^{m \to n}_{(i,j)}(t, \Delta t) \\
\end{split}
\end{equation}
The \(T^{m \rightarrow n}_{(i,j)}\) denotes the probability of the \(i\)th node transitioning from state \(m\) to \(n\) given the network state \(X(t)\), based on its interaction with node \(j\). \(I_{(i,j)}(t)\) is the indicator function, i.e., it is 1 when \(i\) and \(j\) are in contact and 0 otherwise. These interactions are stochastically independent, i.e., the interaction between \(i\) and \(j\) is independent of its interactions with other nodes. Each transition is a Poisson process with rate \(\lambda^{m \rightarrow n}_{i, j}(t)\), therefore the sum of them is also a Poisson process, and we have:

\begin{equation}
\begin{split}
\Pr[x_i(t + \Delta t) &= e_n \mid x_i(t) = e_m, X(t)]  \\
&= \Delta t \sum_{j=1}^{N} \lambda^{m \rightarrow n}_{(i,j)}(t)  + o\Delta t  
\end{split}
\end{equation}

\(\lambda^{m\rightarrow n}_{i, j}\) depends on the node-based transition ($i=j$) and edge -based transition ($i\neq j$) graphs such as:
\begin{equation} 
\lambda^{m\rightarrow n}_{i, j}(t)=
\begin{cases}
 \delta_{m,n}, \hspace{25mm}i=j &\\ 
\sum_{l=1}^{L} \beta_{l,mn} y_{l,i}(t) \hspace{8mm} i\neq j
\end{cases}
\end{equation}
where: 
\[y_{l,i}(t) \triangleq \sum_{j=1}^{N} I_{l,i,j} I_{\{ x_j(t)=e_{q_l} \}} \]
where $y_{l,i}(t)$  is the number of inducers( i.e. $x_j(t)=e_{q_l}$) in contact with node $i$ in $l$th layer at given time $t$,   and $\delta_{mn} \ge 0$  is the rate for a vertex to make  a node-transition, which is independent of the neighbors and it is nonzero only when a possible nodal transition \( m \rightarrow n \) exists. Similarly, \( \beta_{l, mn} \ge 0 \) is the rate for for making an edge-based transition,  \( m \rightarrow n \), based on the  of node $i$'s interaction with neighbors in layer $l$

The interarrival time, also known as the waiting time or jump time between transitions, for each node $i$ in state $s$ is initially exponentially distributed with rate $\lambda_i^s$, where $\lambda^i_s$ is sum of all transition rates possible from state $s$ (\( \lambda^i_s = \sum _{m \in \{1,2,...,M\} } \lambda^{s \rightarrow m}_i)\). Each  time when a transition occurs, the network enters a new state, potentially changing the rates for all nodes. The waiting time until the next transition in the entire network is determined by the node having the minimum of these exponential distributions across all nodes:
\begin{equation}
i = arg\,min \{\tau_i\}, \quad \text{where} \quad \tau_i \sim \text{Exp}(\lambda_i^s)
\end{equation}
This process repeats after each transition,  creating a series of Poisson processes with potentially different rates at each step. 

Therefore, from Equations 2, 3, and 6, it is evident that unlike nodal transitions, edge-based transitions of node \( i \) are conditioned on the states of neighboring nodes within the contact networks, leading to a dependence of the distribution of node \( i \)'s state on the states of other nodes, and thus the equation will not be closed. Furthermore, the joint state of the system defined in Equation 3 as \( X(t) = [x_1(t), \dots, x_n(t)] \) is a Continuous Time Markov Chain (CTMC) with \( M^N \) possible network states, suggesting that the probability of node \( i \) being in any state \( \{1, \dots, M\} \) can be calculated as the marginal distribution of \( X(t) \). requiring, the analytical solution to solve the exact Kolmogorov equations for CTMC, which is nearly insoluble due to the state explosion phenomenon, even for a small number of nodes. Consequently, to avoid using approximations such as moment closure techniques , this work present an exact event-based simulation algorithm which can be scaled to large sized multilayer networks.

In the next part, the proposed method for stochastic simulation based on GEMF will be thoroughly investigated.

\subsection{ Efficient Event Based Simulator: FastGEMF }

\begin{figure}[H]
	\centering
	\includegraphics[width=.7\linewidth]{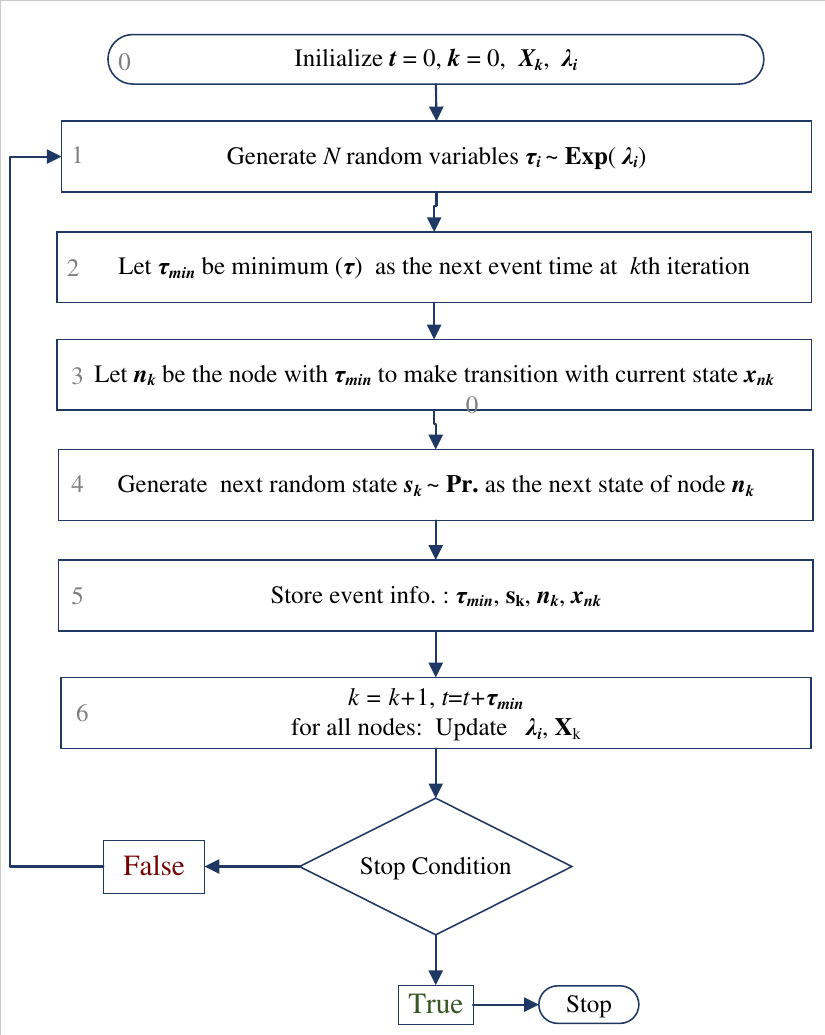}
    \caption{Diagram of conventional  stochastic simulator algorithm modified for GEMF}
\end{figure}

GEMF is based on stochastic network models that explore agent-based interactions within a network of units, focusing on the structure of the contact network rather than the individual attributes of the units. This approach allows for the modeling of complex disease propagation dynamics across diverse network topologies, effectively capturing the heterogeneity inherent in the contact process itself while considering multiple spreads. These types of modeling are particularly valuable due to their ability to simulate a range of outcomes under various epidemic scenarios, driven by their stochastic nature. However, a significant limitation is the computational load required to conduct these simulations, particularly in expansive network models. This computational challenge is a common issue in network-based epidemic modeling, where the efficiency of the simulation algorithms, such as those inspired by the Gillespie stochastic method, becomes crucial in managing the scalability of model implementations. The conventional stochastic simulator algorithm, which is modified for the framework defined by GEMF, can be summarized in Figure 2. In step 0, \( \lambda(k) \in \mathbb{R}^{N \times 1} \) is a vector containing rates for any possible transition of all nodes in the networks at a given moment \( k \), i.e., \( \lambda(k) = \begin{bmatrix} \lambda_1(k), \dots, \lambda_N(k) \end{bmatrix}^T \), and \( x_i \in \{1, \dots, M\} \) is the current state of node \( i \). \( \lambda_i(k) \) can be formulated as a generalization of equation 3, so we have:

\begin{equation}
\begin{split}
&\lambda_i(k)=\sum_{s=1}^{M}A_\delta(x_i(k),s_k) \\+ 
& \sum_{l=1}^{L} \Big( \sum_{s=1}^{M}A_\beta (x_i(k),s_k;l) \sum_{j=1}^{N}w(j,i;l)I_{ \{x_j(k),q(l)\}  } \Big )
\end{split}
\end{equation}

Where \( s_k \) denote the state of a node at moment \( k \), and let \( q(l) \) denote the influencer indices of layer \( l \) and $w(j,i;l) \ge 0$ denote the weight of link between node $i$ and $j$. $A_{\delta}$ and $A_{\beta}$  correspond to adjacency matrices for  the node-based and edge-based transition rate graphs, respectively as we define them  equation 8 and  Figure 3.\
\begin{equation}
\begin{split}
A_\delta=
\begin{bmatrix}
\delta_{11}      & \cdots & \delta_{1M}      \\  \vdots & \ddots & \vdots \\ 
 \delta_{M1}      & \cdots & \delta_{MM}
\end{bmatrix}_{M\times M},
\end{split}
\end{equation}
\begin{figure}[H]
	\centering
	\includegraphics[width=.5\linewidth]{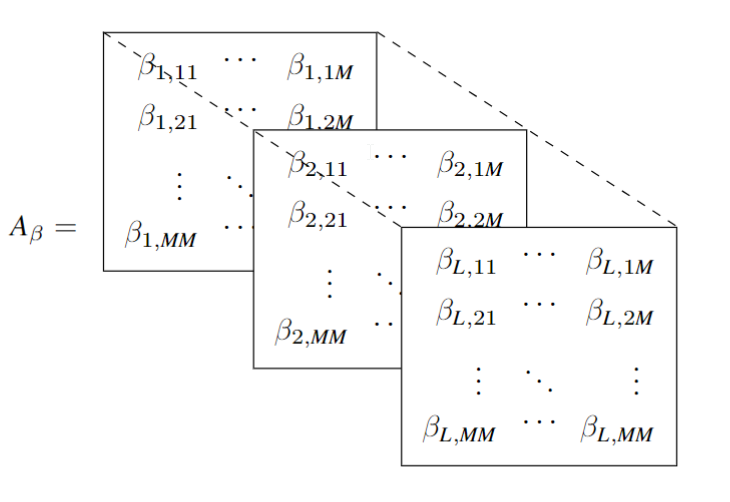}
    \caption{edge -based transition rate matrix}
\end{figure}
Let \( A_\delta(m, n)\ge 0 \) correspond to nodal transition rate from state $m$ to $n$ , and  similarly, \( A_\beta(l,a,b) \ge 0 \)  denote the edge-based transition rate from state  $a$ to $b$ at layer $l$ , where $m, n, a, b \in \{1, 2,\cdots, M\}$ and $l \in \{1,2, \cdots, L\}$.\\
As can be seen from steps 1 to 5 in Figure 3, the process requires the generation of \( \tau \in \mathbb{R}^{N \times 1} \), finding its minimum, and sampling a random variable (r.v.) from the probability distributions \( Pr \in \mathbb{R}^{M \times 1} \), and updating \( \lambda \in \mathbb{R}^{N \times 1} \). These operations result in three operations with linear time complexity , leading to \( \textbf{O}(N) \), which makes the algorithm slow and inefficient even in small network sizes. In \cite{sahneh2017gemfsim}, Sahneh et al. showed that step 1 can be replaced as follows:

\begin{equation}
\begin{split}
\text{find min } \tau_i \sim \text{Exp}(\lambda_i) \text{ for } i = 1, \ldots, N \\
\text{is equivalent to} \\
\tau_{\text{min}} \sim \text{Exp}\left(\sum_{i=1}^N \lambda_i \right) \Rightarrow \tau_{min}=\frac{-ln(u)}{\sum_{i=1}^N \lambda_i}
\end{split}
\end{equation}
Let \( u \) be a random variable (r.v.) with a uniform distribution (\( u \sim \text{Unif}(0,1) \)). It states that the minimum of a set of exponentially distributed random variables itself follows an exponential distribution with a rate equal to the sum of all rates. They also limit the number of updates of \( \lambda_i \) to nodes that are connected to the node that made the transition in event \( k \). These modifications reduce the time complexity to only one linear operation to create the probability distribution for nodes' rates and logarithmic complexity for sampling from that distribution, which still leading to a complexity of \( \textbf{O}(N) \). Therefore, the modified version is also inefficient, having linear time complexity with respect to \( N \). Consequently, we propose the following event-based algorithm as shown in Figure 4.

In our proposed algorithm, we shift focus from relative transition times to absolute times \cite{gibson2000efficient}, enabling the identification of transitions based on their occurrence time. We replaced the need to sample a node having the minimum interarrival time in the network---which has a time complexity of \( \textbf{O}(N) \)---with a data structure that prioritizes  absolute transition times of nodes in non-descending order, reducing the time complexity of finding the minimum time and the node that will make the transition  to constant \(\textbf{O}(1) \). Therefore, to avoid any confusion, we exactly define  what does the absolute time mean and how it is equivalent to interarrival time. So we have definition 1 as:\\
\textbf{Definition 1:} Suppose we have random variables $\tau_i \sim Exp(\lambda_i)$ as the interarrival times for $1<i<N$ as N being the number of nodes and  with PDF \(f_{\tau_i}(u;\lambda)=H(u)\lambda_i e^{ (-\lambda_i u ) }\), where \( H(u) \) is Heaviside function, and is  0  for  $u<0$ and  1 for $ u \ge 0$ . Therefore, we  have the absolute time as a random variable $ T_i=\tau_i+t_{current}$ as sum of interarrival time and $t_{current}$ which is the time that has already passed.\\
Now from definition 1 and using random variable transformation theorem \cite{gillespie1991markov} , we can  derive the PDF of absolute time as:
\begin{equation}
\begin{split}
     f_{T_i}(u)=\int_{-\infty}^{\infty} f_{\tau_i}(u')\delta(u-(u'+t_{current}))du'=\\ 
     f_{\tau_i}(u-t_{current})=H(u-t_{current})\lambda_i e^{(-\lambda_i (u -t_{current} ))}
\end{split}
\end{equation}
Now we can simply show that $T_i$ and $\tau_i$ are equivalent to each other. So first, we define  the memoryless property of the exponential distribution is as the  Theorem 1.\\
\textbf{Theorem 1}: Let T be an exponentially distributed random variable with rate $\alpha$
 (i.e.,  \(T \sim Exp(\alpha) \) ). Then, for any  $s$,$ t$ $\geq 0$, we can say:
\[P(T > s + t \mid T > s) = P(T > t).\]
Now, by writing the CDF for absolute time in in equation 10  as:
\begin{equation}
\Pr(T_{i} > u) = \begin{cases} 
e^{-\lambda_{i}(u - t_{current})} & \text{if } u > t_{current} \\
1 & \text{otherwise} 
\end{cases}
\end{equation}
It is immediately clear from theorem 1  that both relative time and absolute time are equivalent. When $ u < t_{current}$, the transition has already occurred (probability equals 1). When $u> t_{current}$, $ u - t_{current}$ represents the interval in which the next transition will happen (defining the interarrival time). The memoryless property in Theorem 1 proves this interval's independence from the elapsed time $t_{current}$, thus establishing the equivalence of absolute times generated in step 6 of Figure 4 with relative times generated in step 1.\\
This modification eliminates the need to generate a set of random variables $\tau_i$s in each iteration, as shown in Figure 3, or to sample a node every iteration, as required by the previously mentioned inefficient algorithm, GEMFsim. Instead, we initially generate the relative times for all nodes and sort them in non-decreasing order, as shown in step 1 of Figure 4. Then, in each iteration,finding the node with the minimum time is immediate, and then we  update the absolute time only for the node that has undergone transmission and for nodes whose rates have changed during this process, as shown in step 6.
We can summarize this modification as: 
\begin{itemize}
\item The substitution of relative time ($\tau_i \sim \text{Exp}(\lambda)$) with absolute time ($T_i \sim t_{\text{current}} + (\tau_i \sim \text{Exp}(\lambda))$), where $T_i$ represents absolute times and $t_{\text{current}}$ is the minimum of the absolute times, denoted as $t_{\text{min}}$.
\item The reuse of generated $T_i$ values for nodes not involved in the transition and whose rates ($\lambda_i$s) did not change.
\end{itemize}

However, the re-use of absolute times might raise concerns regarding the claimed exactness of our method.  This can also be proved by the memoryless property in theorem 1, which confirms that for nodes whose rates remain unchanged, the precomputed absolute times \(T_i\) continue to be accurate. This stems from the fact that the probability distribution of the time until the next event does not depend on the elapsed time but is solely a function of the unchanged rate parameters \(\lambda_i\).\\
Furthermore, a cautious update approach is implemented in step 6 to minimize the calculation of updates made in each iteration. While node $n_k$ that transitions to a new state always updates, this approach imposes three constraints on other network nodes to prevent unnecessary calculations. Node $i$'s rate updates only if:
\begin{enumerate}
    \item   Node $n_k$ is transitioning  $to$ or $from$ an inducing state for that layer($x_{nk}=q_l$  \textit{ or } $s_k=q_l$). For instance, in the Susceptible-Exposed-Infectious-Recovered (SEIR) model, a transition from state $S$ to $E$ does not affect other nodes' states or rates. 
    \item Node $i$ is among the immediate neighbors of the node $n_k$ , $A_{l,ik} \ne 0$, which A is the adjacency matrix for  network layer $l$.
    \item The rate change for node $i$ is non-zero ($\Delta \lambda_i\ne 0$). This means, the node is only affected  in the immediate neighborhood of the transitioning node, $n_k$, if there exists a possible path in the edge-based transition graph to either the former ($x_{n_k}$) or current state ($s_k$) of node $n_k$.
\end{enumerate}

To make on the third constraint more clear, consider node $i$ as a neighbor that satisfies the first two conditions with current state $x_{i}$. The update will only proceed if either $A_{\beta}(x_{i},x_{n_k})$ or $A_{\beta}(x_{i},s_k)$ is non-zero. To illustrate this approach, Figure 5 presents a pedagogical example. Consider a transition from state $I$ (Infectious) to $R$ (Recovered). In this scenario, only susceptible neighbors require recalculation of transition rates and generation of new absolute times. This cautious update strategy significantly reduces unnecessary computations, especially in network models where specific state transitions affect only neighboring nodes. These constraints ensure efficient allocation of computational resources, focusing on updates that meaningfully impact the network's stochastic dynamics.\\

These configurations reduce the time complexity to approximately (\textbf{O}($D_{max}\log(N))$), N being number of nodes and $D_{max}$ maximum node degree over all layers. In the next section, we demonstrate that using optimized data structures and coding techniques, this logarithmic time complexity can be approximated as constant time, independent of the number of influencers ($q_l$), nodes, or layers  in the network and linear with the maximum node degree across layers.

\begin{figure}[H]
	\centering
	\includegraphics[width=.6\linewidth]{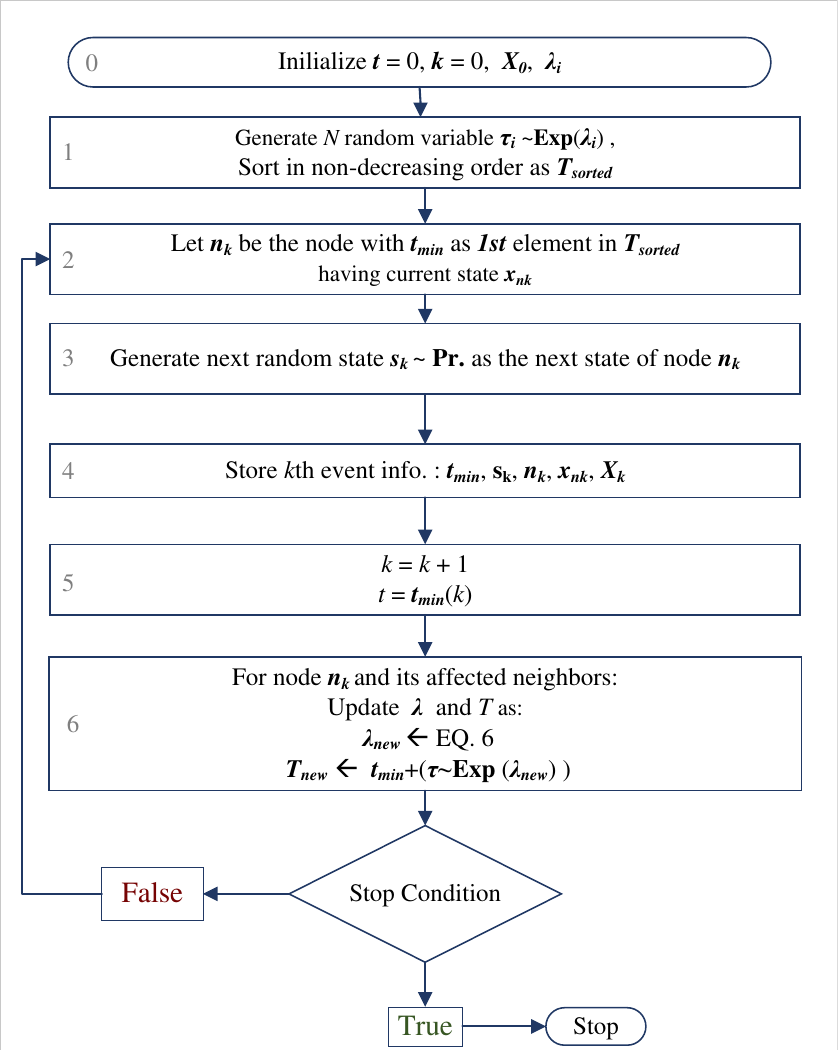}
    \caption{ proposed  optimized stochastic simulator algorithm}
\end{figure}
\begin{figure}[H]
	\centering
	\includegraphics[width=.5\linewidth]{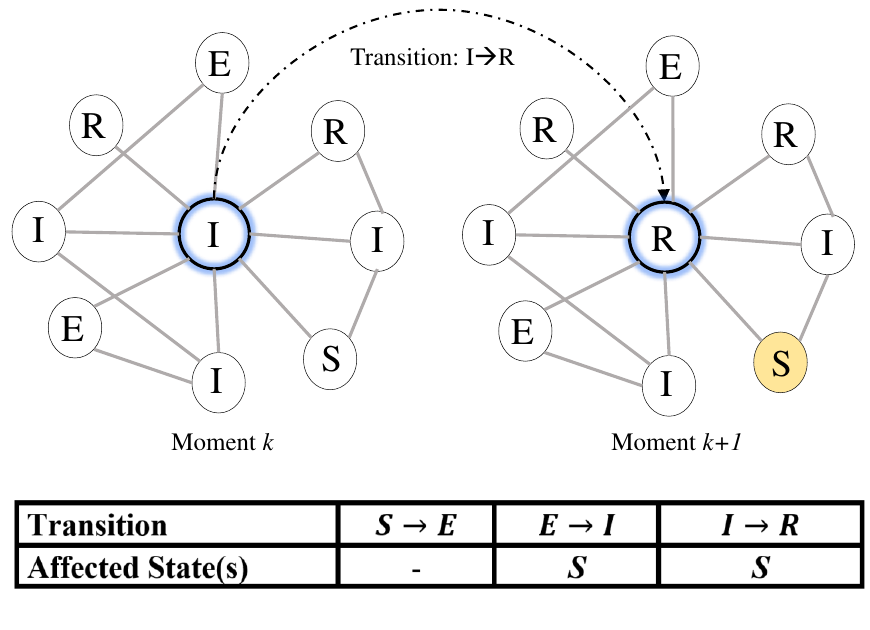}
    \caption{An example  of cautious update approach, demonstrates a state transition from Infected (I) to Recovered (R) in  SEIR model. The highlighted yellow color represents the only nodes undergoing update at this timestep. Below the graph, the transition effect matrix  specifies the states affected by each transition}
\end{figure}

\subsection{Implementation}
In this section, we will briefly discuss the libraries and data structures utilized in FastGEMF. Additionally, we will present a detailed pseudo code to provide a comprehensive, step-by-step framework for this research.\\

Given the inefficiency of using an adjacency matrix for storing network structure due to its \(\textbf{O}(N^2)\) space complexity, we employed Compressed Sparse Row (CSR) format, also known as the Yale format, as well as Compressed Sparse Column (CSC) format. These sparse matrix representations are significantly efficient for both storage and computation, particularly in large-scale networks.Using both   CSR and CSC  in FastGEMF network structure, lead to optimized memory usage and enhance computational performance  and  thereby reducing the overhead associated with matrix operations, compared to using the well-defined adjacency matrix format. The utilization of these methods also allows for optimized look-up through multilayer networks. As shown in Figure 6, this enables us to efficiently identify nodes that influence their neighbors, as well as nodes influenced by a specific node. This capability allows us to handle different structures of directed and undirected networks within a unified and robust framework without any prior assumption about the structure of layers. We used the SciPy Sparse library in Python for efficient implementation of network structure and look-up methods as described. \\
\begin{figure}[H]
	\centering
	\includegraphics[width=0.5\linewidth]{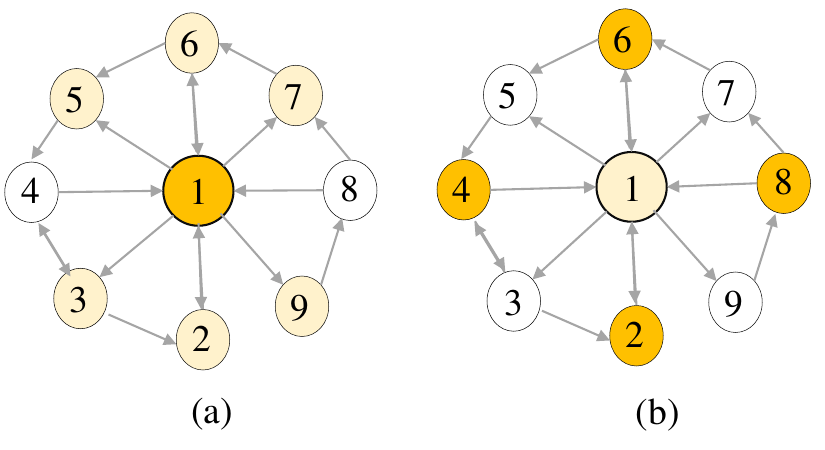}
    \caption{Two types of network search in FastGEMF. (a) Nodes that can be influenced by node 1. (b) Nodes that can influence node 1. The orange and yellow nodes represent the influencing and influenced nodes in each look-up method, respectively. }
\end{figure}

Due to the cautious update approach we implement in our algorithm, typically a few number of nodes' rates($\lambda_i$) and interarrival times ($t$) are updated in every event. Updating the $\lambda$s are efficient and fast due to high speed python NumPy arrays computation we used. However, the $t$s are organized in non-decreasing  order which make the data structure more complicated, and though finding the minimum of interarrival times is constant time, $\textbf{O}(1)$, the updating
can be more time consuming. There are different ways for constructing an indexed priority queue, but they are mostly designed to be efficient for only finding the highest priority (pop) and adding new values (push), which are not efficient for GEMF algorithm as each event along finding the minimum, specific nodes need to be updated.Various types of heaps (e.g., binary, skew, binomial, etc.) are among most famous examples of these structures. For efficient update of $\tau$ we used the sorted list structure by \textit{SortedContainer} module in Python. This data structure takes advantage of fast operations of python's \textit{list}, offering an efficient algorithm for sorting the list using smaller fragments for sorted containers and storing them in lists of lists, which  allows  updates and modification of all members with time complexity which theoretically is \( \textbf{O}(\log(N))\), which is  the  efficient data structure for FastGEMF algorithm compared to conventional tree-based implemented ones . More details about this data structure can be found at \cite{sortedcontainers}.\\

Large scale networks usually have sparse structures where  a few number of connections exist between agents compared to size of network.However, to maintain the scalability of the algorithm with respect to number of edges in network, we also implement the same algorithm with a little modification such that for dense contact layers we used the NumPy array to store $\tau$ and finding the minimum. this approach is trade off between using NumPy arrays having  minimum finding time  complexity of  $\textbf{O}(N)$) and  updating of ($\textbf{O}(D_{max})$) and sortedlist where finding the minimum is  constant time $\textbf{O}(1)$ and updating  is $\textbf{O} (D_{max}\log(N))$. The FastGEMF  automatically choose which one to use based on classifier trained with alot of benchmarks for different network sizes. The implementation of FastGEMF is summarized in algorithm 1.Furthermore, As a common practice in numerical computing, the inherent inefficiency of Python loops has been avoided through the extensive use of vectorized calculations in NumPy, leveraging its highly optimized performance characteristics.
\begin{algorithm}[H]
\caption{FastGEMF algorithm}\label{Alg-Decap}
\begin{spacing}{1.2}
\begin{algorithmic}[1]
  \Require $\mathcal{G} (\mathcal{N},E_1,E_2,\dots,E_L), A_\beta, q, A_\delta, X_0$, \textit{Stop Condition} 
  \Ensure Output event
  \State \textbf{Initialization:}
  \State net $\leftarrow$ Network($\mathcal{G}$) \hspace{5mm}\(\triangleright\) create network object
  \State $X \leftarrow X_0$
  \State $\tau_{sorted} \leftarrow$ Tsorted([]) \hspace{5mm} \(\triangleright\) create empty sorted list object
  \State $\lambda \leftarrow \mathbf{0}_{N \times 1}$
  \State $n_{inf} \leftarrow (X == q).indices$ \hspace{5mm} \(\triangleright\) agents being initially influenced
  \State $n_{aff} \leftarrow$ net.get\_out\_neighbors($n_{inf}$) \hspace{5mm} \(\triangleright\) affected agents
  \ForAll{$l$ in $L$}  
  \ForAll{$n$ in $n_{inf}$ \textbf{or} $n_{aff}$}
      \State $W_{nbr}(n, l) \leftarrow$ net.get\_in\_weights($l, n$) $\times I(X[n], q)$
    \EndFor
    \State $\lambda_n \leftarrow \sum_{s=1}^{M} A_\delta(X[n], s) + \sum_{l=1}^{L} A_\beta(X[n], s; l) \times W_{nbr}(n,l)$
    \State $\tau_n \leftarrow t \sim \text{exp}(\lambda_n)$
    \State $\tau_{sorted}$.add($\tau_n$)
  \EndFor

  \State $k \leftarrow 0$
  \While{\textbf{not} Stop Condition}
    \State $(n_k, t_k) \leftarrow \tau_{sorted}.\text{pop}()$ \hspace{5mm} \(\triangleright\) agent which makes transition
    \State $x_{n_k} \leftarrow X[n_k]$
    \ForAll{$s$ in $M$}
      \State $\text{rate}(x_{n_k}, s) \leftarrow A_\delta(x_{n_k}, s) + \sum_{l=1}^{L} A_\beta(x_{n_k}, s; l) \times W_{nbr}$
    \EndFor
    \State $Pr \leftarrow \frac{\text{rate}}{\sum_{s=1}^{M} \text{rate}(x_{n_k}, s)}$
    \State $x_{n_k}^{\text{new}} \sim Pr$  \hspace{15mm}  \(\triangleright\) sampling next state $x_{n_K}^{new}$ from prob. dist. $Pr.$
    \State $X[n_k] \leftarrow x_{n_k}^{\text{new}}$
    \State $\lambda(n_k) \leftarrow \sum_{s=1}^{M} A_\delta(x_{n_k}^{\text{new}}, s) + \sum_{l=1}^{L} A_\beta(x_{n_k}^{\text{new}}, s; l) \times W_{nbr}(n,l)$
    \If{$\lambda(n_k) \ne 0$}
      \State $\tau_{n_k} \leftarrow t_{k} + (t \sim \text{exp}(\lambda(n_k)))$ \hspace{5mm}  \(\triangleright\) sample new event time for  $n_k$
      \State $\tau_{sorted}$.add($n_k, \tau_{n_k}$)
    \EndIf
    \State event $\leftarrow (t_k, n_k, x_{n_k}, x_{n_k}^{\text{new}})$
    \ForAll{$l \mid (q(l) == x_{n_k} \text{ or } q(l) == x_{n_k}^{\text{new}})$}  \hspace{5mm}\(\triangleright\)   cautious updating 
      \State $\delta \leftarrow I(q(l), x_{n_k}^{\text{new}}) - I(q(l), x_{n_k})$
      \State $n_{aff} \leftarrow$ net.get\_out\_neighbors($n_k, l$)
      \ForAll{$n$ in $n_{aff} \mid (A_\beta(x_n, x_{n_k}; l) \ne 0, A_\beta(x_n, x_{n_k}^{\text{new}}; l) \ne 0)$}
        \State $W_{nbr}(n, l) \leftarrow W_{nbr}(n, l) + \delta \times$ net.get\_in\_weights($n, n_k; l$)
        \State $\lambda(n) \leftarrow \lambda(n) + \delta \times \sum_{s=1}^{M} W_{nbr} A_\beta(x_n, s; l)$
        \If{$\lambda(n)\ne 0$}
            \State $\tau_{n} \leftarrow t_{k} + (t \sim \text{exp}(\lambda(n)))$
            \State $\tau_{sorted}.update(n,\tau(n))$
        \Else
            \State $\tau_{sorted}.remove(n,\tau(n))$
        \EndIf   
      \EndFor
    \EndFor
    \State $k \leftarrow k + 1$
    \State Update Stop Condition
  \EndWhile
\end{algorithmic}
\end{spacing}
\end{algorithm}

\section{Performance of FastGEMF}
To evaluate the efficiency and scalability of the proposed algorithm, we conducted a series of benchmarks comparing its performance against GEMFsim, the well-known simulator Network Diffusion Library, NDlib\cite{rossetti2018ndlib}, which is being constantly updated since its release, and newly released module 'epidemik'\footnote{https://github.com/DataForScience/epidemik/}, all of which are in Python language and can simulate multi-influencer  compartmental models, however, NDlib and epidemik  only simulate spread process over single layer and unweighted networks. It is important to note that this analysis is not comprehensive due to the different natures and purposes for which each model is designed.\\

NDlib and epidemik  utilize the NetworkX library to handle network structures and look-up through network, which limits handling operations over network with different layers. To ensure a fair comparison and minimize the impact of the underlying libraries on performance, we used the same single-layer unweighted-undirected contact network across all simulations.
Given that NDlib simulates the spreading process using a discrete-time approach, we adopted a simple SIR model and compared the time required to simulate the process until no infected individuals remained and probabilities of discrete approach for the SIR model are regulated for CTMC rates as:
\begin{equation}
    \begin{aligned}
    \beta = \frac{-\ln(1 - P(\text{S} \to \text{I}))}{ \Delta t}\\
    \delta = \frac{-\ln(1 - P(\text{I} \to \text{R}))}{\Delta t}
    \end{aligned}
\end{equation}
while $\beta$ and $\delta$ being the infection and recovery rates, respectively and \(P(\text{S} \to \text{I})\) and \(P(\text{I} \to \text{R})\) are probability of infection and recovery transmissions.  It is worth mentioning that the discrete-time approach approximates continuous-time processes, and with a sufficient number of simulations and suiTable time steps, the average results and probability distribution of the stochastic process will converge to those of continuous-time simulations. As shown in Figure 8, the average fraction of states for over 3000 simulations of SIR model over a same contact network presented for GEMFsim and FastGEMF as CTMC-based simulators and NDlib as DTMC-based simulator of the spread process. \\
\begin{figure}[H] 
	\centering
	\includegraphics[width=1\linewidth]{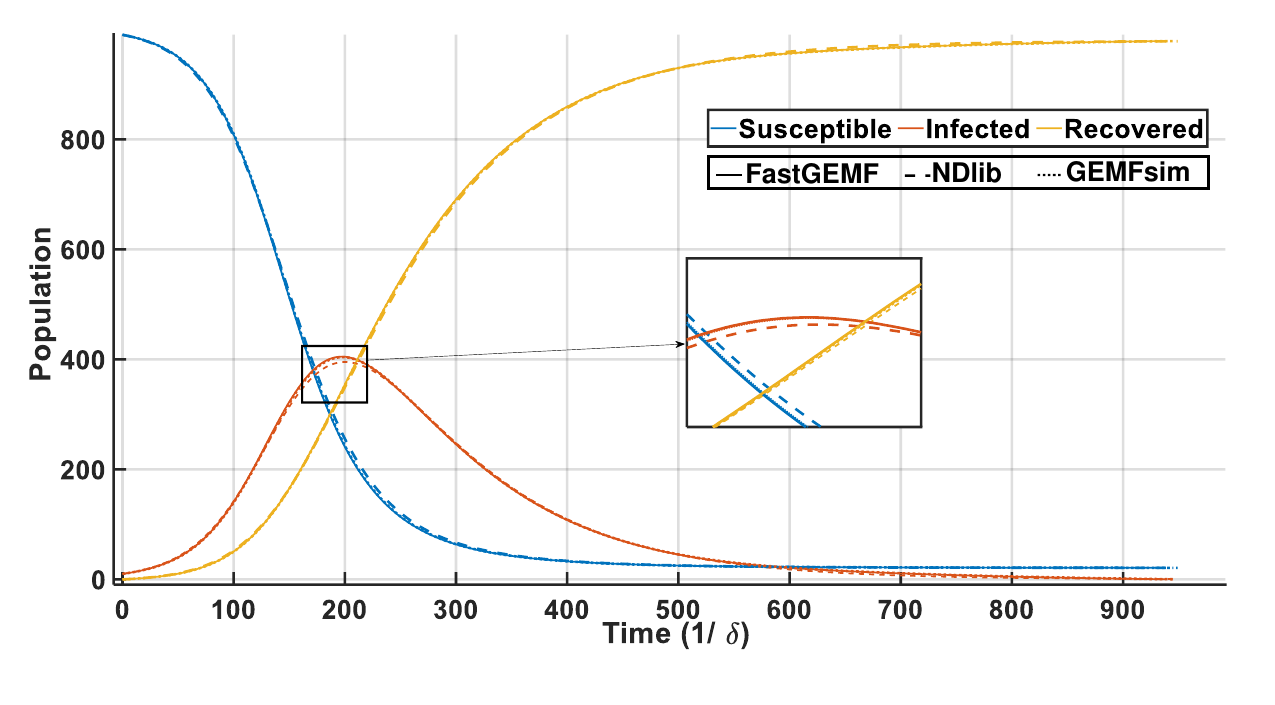}
	\caption{Average Population for SIR model over 3000 simulation on  Erd\H{o}s--R\'{e}nyi
 network with 1000 nodes  with probability of p=0.2 and model rates as $\beta$=0.0002 and $\delta=0.01$ and step-size 1 in NDlib.}
 \end{figure}
 All of the following benchmarks were executed by a machine having 256 GB RAM and two processors of Intel(R) Xeon(R) CPU X8260 @2.40 GHz. 
 
 \renewcommand{\arraystretch}{1.5}
\begin{table}[b]
\centering
\begin{tabular}{|l|c|c|c|c|c|}
\hline
\multirow{2}{*}{\textbf{Simulator}} & \multicolumn{5}{c|}{\textbf{Contact Network Size}} \\ \cline{2-6}
 & \textbf{10\textsuperscript{2}} & \textbf{10\textsuperscript{3}} & \textbf{10\textsuperscript{4}} & \textbf{10\textsuperscript{5}} & \textbf{10\textsuperscript{6}} \\ \hline
fast\_SIR  &0 .0073s  & 0.03s & 0.41s & 6.29s   & 95.89s\\ \hline
FastGEMF & 0.023s  & 0.22s & 1.75s & 25.29s   & 378.12s \\ \hline
NDlib & 0.188s      & 1.57s & 16.6s & 206.23s & 2956.74s \\ \hline
epidemik & 0.457s  & 4.71s & 49.6s & 551.7s   & 5860.1s \\ \hline
GEMFsim & 0.05s    & 0.59s & 12.32s & 787.45s & Nan \\ \hline
\end{tabular}
\caption{Running Time Comparison For SIR Simulation: results for   random geometric network with different sizes while the average the degree was kept constant about 11. SIR model rates are as$\beta$=0.005 and $\delta=0.01$  }
\label{Table:simulation_results}
\end{table}
To investigate the scalability of the modules with respect to the number of nodes in the network, we conducted a series of simulations using the SIR model, varying the network size from 100 to 1 million nodes, having random geometric structure with average edge degree of 11. The results of this benchmark are detailed in Table 1. We also include the fast\_SIR from Epidemics on Network (EoN) library \cite{kiss2017mathematics}, which is a specialized implementation of the SIR model intended to be scalable on single-layer networks. fast\_SIR assumes that the duration of contact between an infected agent and a susceptible one does not matter; the only condition is that the agent was infected at initial time of interaction and be infected at the time the susceptible neighbor transitioned to infected under its influence. Therefore, the possible transitions of influencer node that might happen in this interval does not matter.The comparison between FastGEMF and fast\_SIR is meaningful as it contrasts two extreme modules: FastGEMF as a general and flexible module, and fast\_SIR as a highly specialized one. As shown in Table 1, fast\_SIR is approximately  4 times faster than FastGEMF, which can be considered an accepTable trade-off for our module given its generality and flexibility.Considering the event as single nodal or edge-based transition, the event time of the benchmarks are showed in Figure 9. The performance analysis reveals that FastGEMF demonstrates superior scalability and performance compared to the other methods. This is particularly evident as the network size approaches 1 million nodes. As shown in Figure 9, when the network size increases by a factor of \(10^4\) from 100 to 1 million, the event time for FastGEMF increases by approximately 71\%. In contrast, the predecessor GEMFsim exhibits a linear increase in event time with network size.

 While NDlib and epidemik shows  at least 10 times slower performance, they exhibit similar time complexity to FastGEMF (almost constant with respect to  number of nodes). That's because the simplifying assumption of networks to be unweighted. In this context, the probability of edge-based transition from one state to another in an unweighted network is uniform across all edges. This results in a constant time sampling complexity of \textbf{O}(1). However, in weighted networks,  possible combinations of infected nodes and susceptible  neighbors and their corresponding weights must be considered, leading to increased computational complexity. Specifically, the time complexity for creating the distribution is typically linear, and the complexity for sampling from it is logarithmic. This significantly increases the computational time required for each iteration. As a result, DTMC simulators often approximate networks as unweighted to avoid the extensive calculations associated with weighted networks. However, In FastGEMF, the weight of the network does not impact the running time.

 \begin{figure}[H][t]
	\centering
	\includegraphics[width=.75\linewidth]{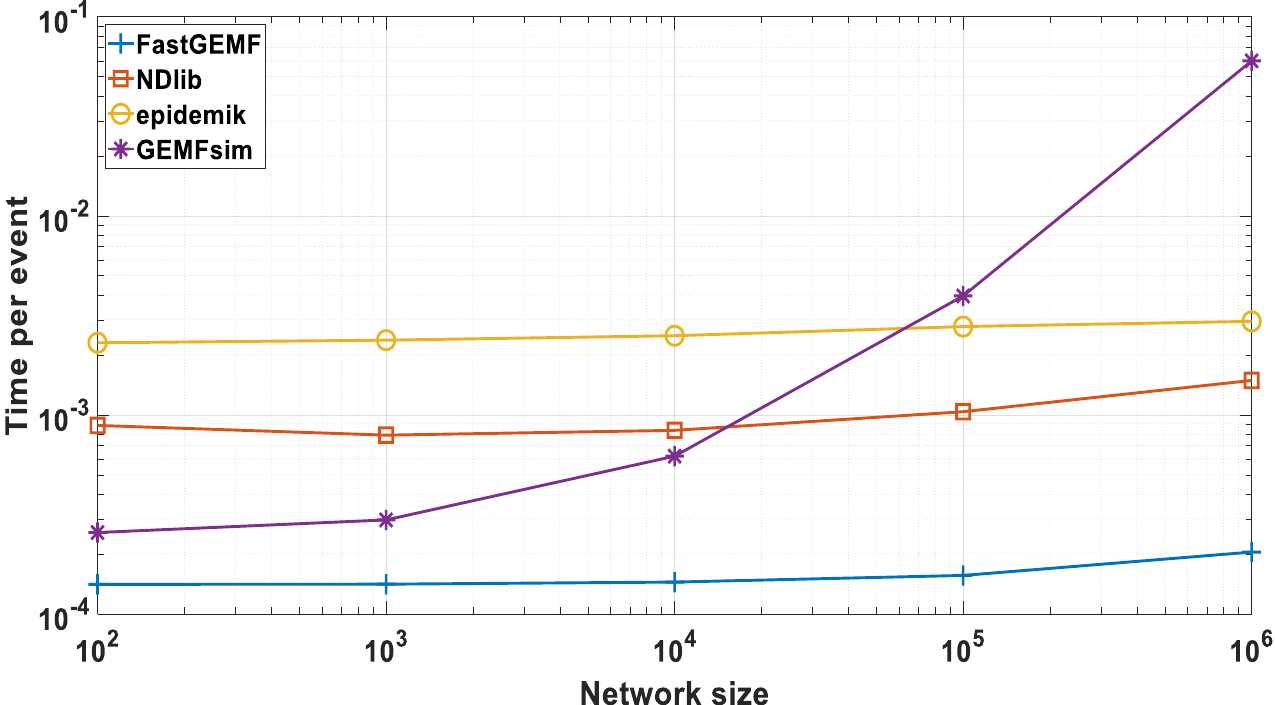}
	\caption{increasing  network size  for  from $10^2$ to $10^6$ with constant node degree equals 10 in undeirected random geometric network.}
 \end{figure}

 Furthermore, we tested the performance of different methods over scaling by increasing the number of edges for a network of 5,000 nodes from 10 links per node(50k edges) to 4,999($\approx 12.5M$ undirected edges) as a complete graph. The results are shown in Figure 10. The superior performance of FastGEMF is again evident, showing the running times remained on the scale of $10^{-4}$ even in a complete graph, while the NDlib and epidemik ranged from almost $10^{-3}$ to  $10^{-2}$ and $10^{-1}$, respectively. However having a complete graph with this size is very rare in real life application, it exhibits that even in very dense networks FastGEMF has a promising performance.\\
\begin{figure}
	\centering
	\includegraphics[width=.75\linewidth]{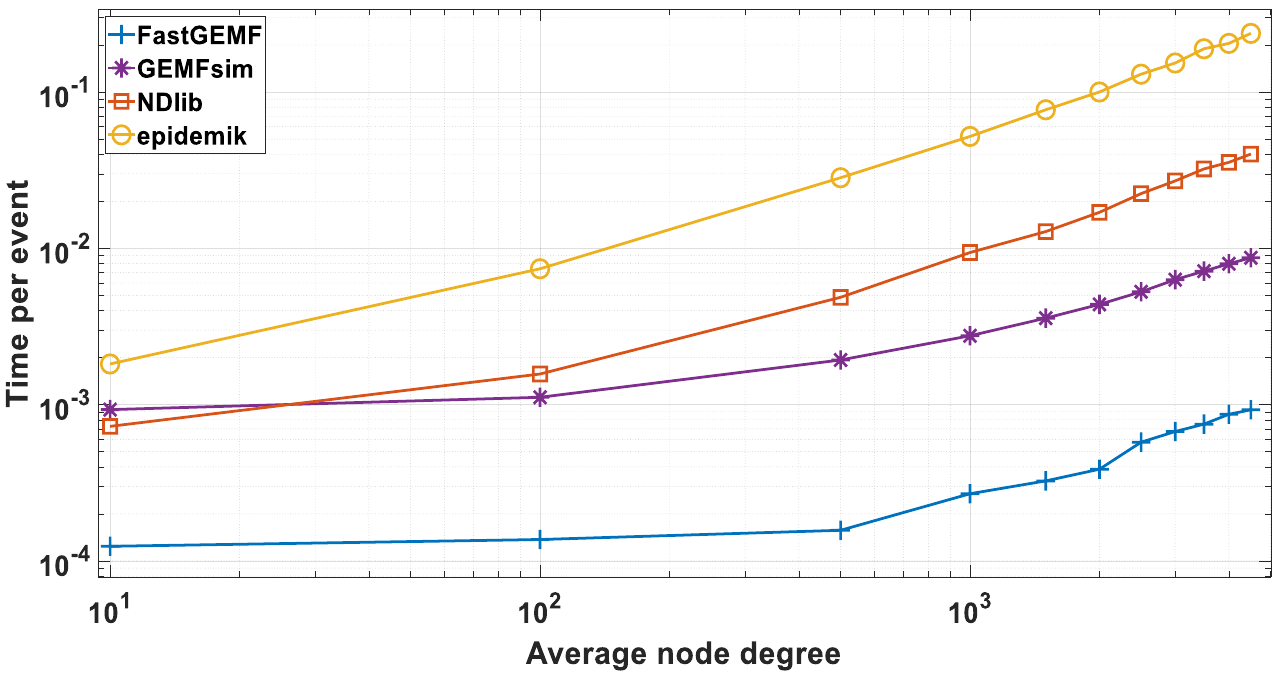}
	\caption{Scaling network edges for a network with 5000 nodes and average node degree from 10 to complete graph.}
 \end{figure}
Finally, in Figure 11, the scalability of the FastGEMF is shown while scaling the networks from 1 to 100 layers for both weighted and unweighted networks. Different network structures such as random geometric, Watts–Strogatz, Erd\H{o}s--R\'{e}nyi, Barbasi-Albert, etc. with different network centrality tested, and all showed that same constant performance. We use the $SI_1I_2\dots I_L S$ model  as extends the basic SIS model by incorporating multiple types of infections that can affect a susceptible node \cite{darabi2014competitive}. In this model, we consider a scenario where L different infections can  simultaneously compete to spread  through the population. Each node in the network can be in one of L+1 states: susceptible (S) or infected by one of the L types of infections (I1, I2, ..., IL).
Similar to the SIS model, infected nodes recover with a rate $\delta$, regardless of the infection type($\delta_1,\delta_2,\cdots=\delta$). However, the infection process is more complex. A susceptible node can transition to any of the L infected states \( (I_1, I_2, ..., I_L) \) through contact with nodes infected by the corresponding infection type. The infection rates may vary for each type, denoted as \(\beta_1, \beta_2, ..., \beta_L \).
To account for the different transmission dynamics of each infection type, we assume that they spread and compete through distinct contact layers. This means that a susceptible node undergoes a transition to infected state $I_k$ with a rate $\beta_k$ if it is in contact with an $I_k$ node through the kth layer of contact. Figure 12 shows the  transition graph for general $SI_1I_2\dots I_L S$ model.
It is evident that scaling layers with $SI_1I_2\dots I_L S$ does not have a significant effect on the event-time of  FastGEMF as expected. It is noteworthy to mention these event times are only simulation time, without considering initialization (Obviously, the initialization time will be different which is only  one-time calculation ).More benchmarks such as spread over real-world networks, Pokec Social media, Facebook-like social network \cite{leskovec2016snap}, can be be found at FastGEMF documentation. Furthermore, some of capabilities of this framework which are now available at salable level are such as prediction of phase transition and spill-over modelling at inter-connected networks\cite{das2024sir}, extinction time \cite{sahneh2017gemfsim}, modeling Vector-Borne disease spread \cite{ferdousi2019understanding}, computer virus spread \cite{muthukumar2024optimal} and some other which studied at \cite{das2024sir, vajdi2020multilayer,
yi2022multilayer,hosseini2024parsimonious }.
Other modules' perfomances, such as  EpiModel (R language), ComplexNetworkSim (Python language), and Nepidemix (Python language), can be found in the NDlib documentation,as  our comparison focused solely on NDlib and epidemik alongside GEMFsim and FastGEMF.

\begin{figure}[H]
    \centering
    \includegraphics[width=0.65\linewidth]{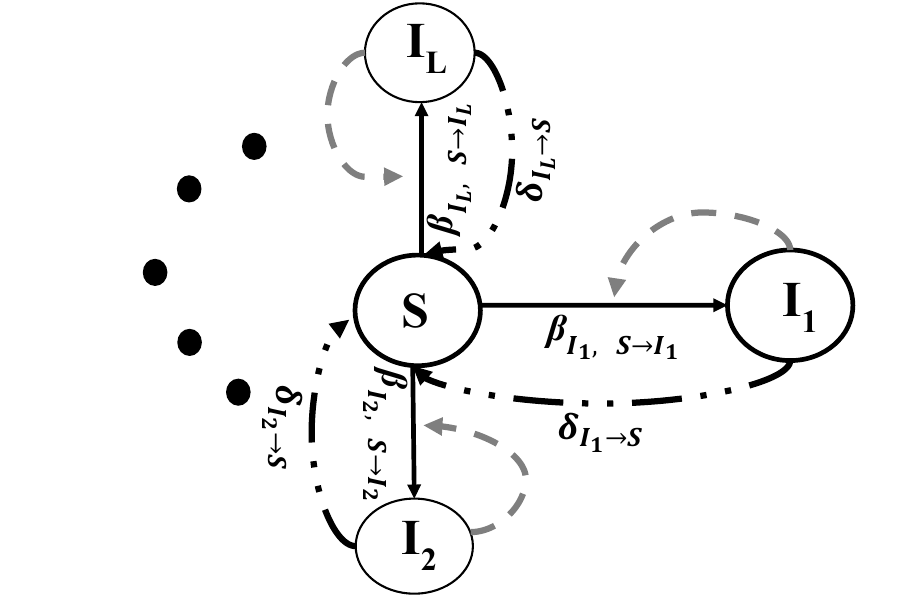}
    \caption{Transition rate graph for $SI_1I_2\dots I_L S$ model used for layer  scaling benchmark, while each layer $k$ has its own inducer $I_k$}
\end{figure}
\begin{figure}[H]
    \centering
    \includegraphics[width=.75\linewidth]{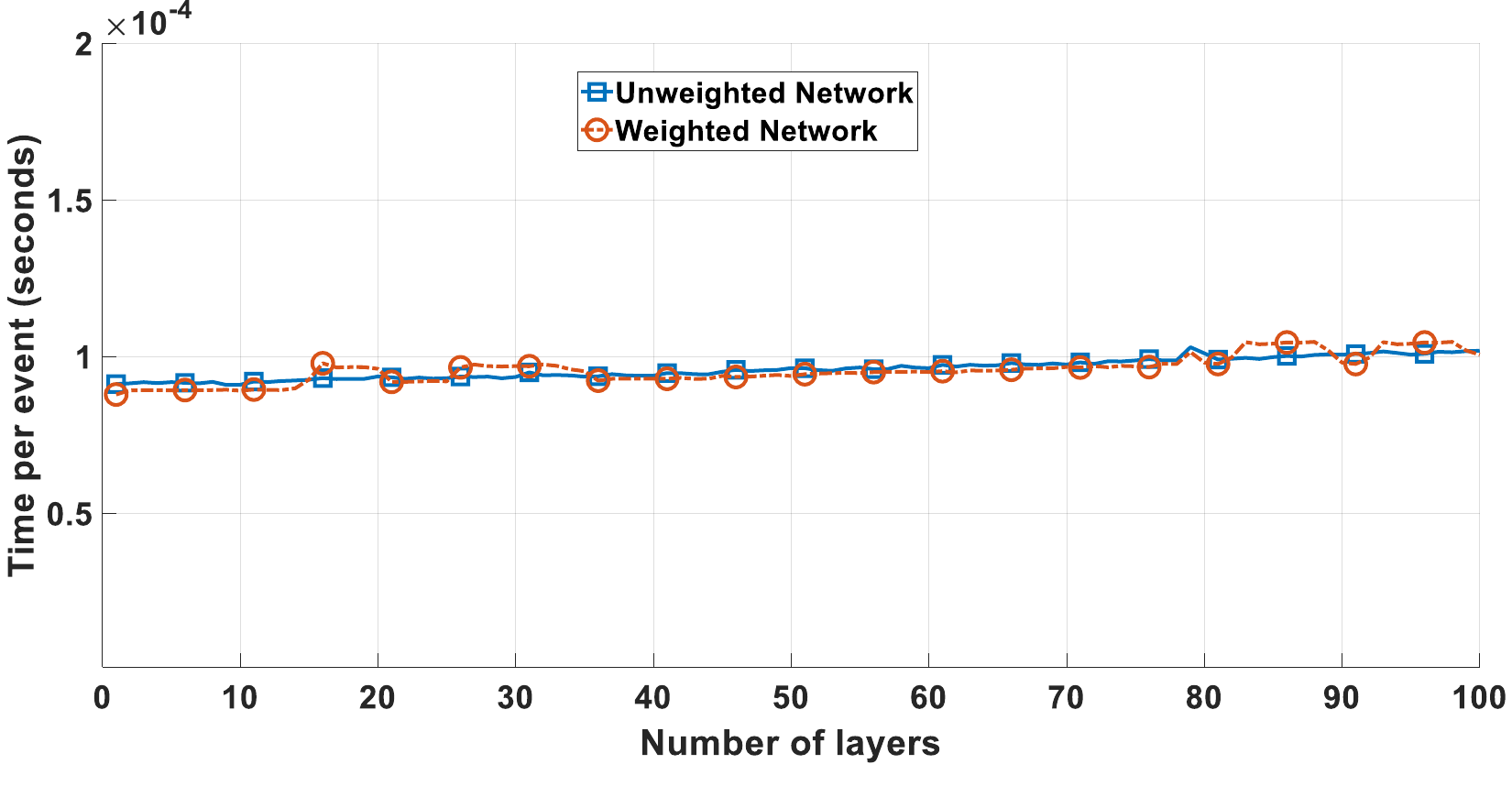}
    \caption{Running time comparison for scaling number of layers from 1 to 100, simulating the spread of $SI_1I_2\dots I_L$ over both unweighted and weighted networks}
\end{figure}

\section{Conclusion}
Spreading phenomena over complex networks have consistently presented challenges due to the intricacies involved in solving the exact stochastic equations of multi-compartmental models, particularly when multiple network layers are involved. This study introduces an efficient and scalable algorithm based on the principles of the Generalized Epidemic Framework(GEMF) to simulate the exact Continuous-Time Markov Chain (CTMC) over complex multilayer networks without resorting to approximations, specializations, or limitations on a number of model states or network structures.\\
The implementation and performance of the proposed framework have been extensively discussed, analyzed, and compared to the existing methodologies. Our results demonstrate the superior scalability and efficiency of FastGEMF with respect to network size, density, and number of layers. Notably, the framework operates without prior assumptions regarding the weighted or directed nature of the graphs.
Implemented as a module in the high-level programming language Python with easy initialization, FastGEMF allows users to define any multi-agent mechanistic models and simulate over networks. This framework facilitates widespread application across diverse domains, including cybersecurity, information technology, epidemiology, and financial markets. Its capacity to elucidate latent spreading phenomena in large-scale real-world networks not only advances our understanding of complex systems but also provides an exact baseline model for researcher for comparative analysis between their methods to investigate the accuracy of their solution.\\

Finally, the authors believe the introduction of FastGEMF represents a significant contribution to the field of network science, offering a versatile tool for investigating stochastic spreading processes on multi-layer networks. As the scientific community continues to explore and refine this framework, it is anticipated to yield valuable insights into the dynamics of complex  systems and foster further advancements in the study of spreading phenomena  for both theoretical research and practical applications.\\

Future work on FastGEMF will focus on three key enhancements: implementing the $\tau$-leap method to accelerate simulations for  large networks with acceptable approximation and incorporating support for dynamical networks. These additions will  boost the framework's efficiency and versatility, enabling more flexible modeling of complex, time-varying systems in different fields of network science.

\subsection{FastGEMF installation and usage }
FastGEMF is now available at Python programming language. For detailed instructions on installing, using, and contribution to FastGEMF, please refer to the official GitHub repository at $https://github.com/KSUNetSE/FastGEMF$ 

\printbibliography
\end{document}